# Observation of Magnetic Superviscosity in Metglas 2605SA1

Anthony C. Crawford    acc52@fnal.gov    Fermilab/Technical Division/SRF    16Nov15

A quasi time constant for magnetic domain relaxation of Metglas 20605SA1 has been measured to be 2.6 hours. The effective relative magnetic permeability has been measured to be 57,000 after relaxation.

## Introduction

Metglas 2605SA1, a low loss transformer core material from Hitachi Metals, is useful as a magnetic shield material for certain applications in superconducting RF cavity cryomodules [1]. In an effort to better understand the shielding performance of this material, measurements have been made of its attenuation of a steady state ambient magnetic field with a magnitude of approximately 500 milliGauss ( $5x10^{-5}$ Tesla ).

In the process of taking measurements it was observed that the field magnitude inside a cylindrical magnetic shield would continue to change for long intervals of time following any manipulation of the position of the shield. Since the ambient field the shield was exposed to was varying little, the change in attenuation could only be attributed to internal changes in the magnetic structure of the Metglas 2605SA1. This phenomenon, called magnetic viscosity, or magnetic relaxation, has been observed and studied previously. A particularly lucid description, with a thorough historical reference list, is to be found in the article authored by Galenko and Branovitskaya [2].

The purpose of this study is to investigate magnetic viscosity effects in Metglas as they relate to shielding DC magnetic field, especially for applications appropriate for superconducting RF cavities. The case of a thin walled cylinder with applied field transverse to the cylinder axis is used. This is a common geometry encountered in shielding superconducting RF cavities.

## Magnetic Viscosity and the Domain Theory of Ferromagnetism

The concept of magnetic domains, also known as “Weiss domains”, has been in existence for more than one century [3]. The subset of domain theory that is thought to be particularly appropriate for slowly varying changes in magnetization within ferromagnetic materials is a re-arrangement of domain boundaries that leads to a reduction of the Gibbs free energy of the system. The shift of interdomain boundaries proceeds at a much slower rate and at lower field than does re-alignment of magnetization vectors within constant domain boundaries. A domain boundary re-arrangement that requires no formation of new boundaries and where intradomain magnetization per unit volume is constant is illustrated in Figure 1. As the size of domains that were pre-aligned with the applied field grows, the total magnetization of the specimen in the direction of the applied field increases.

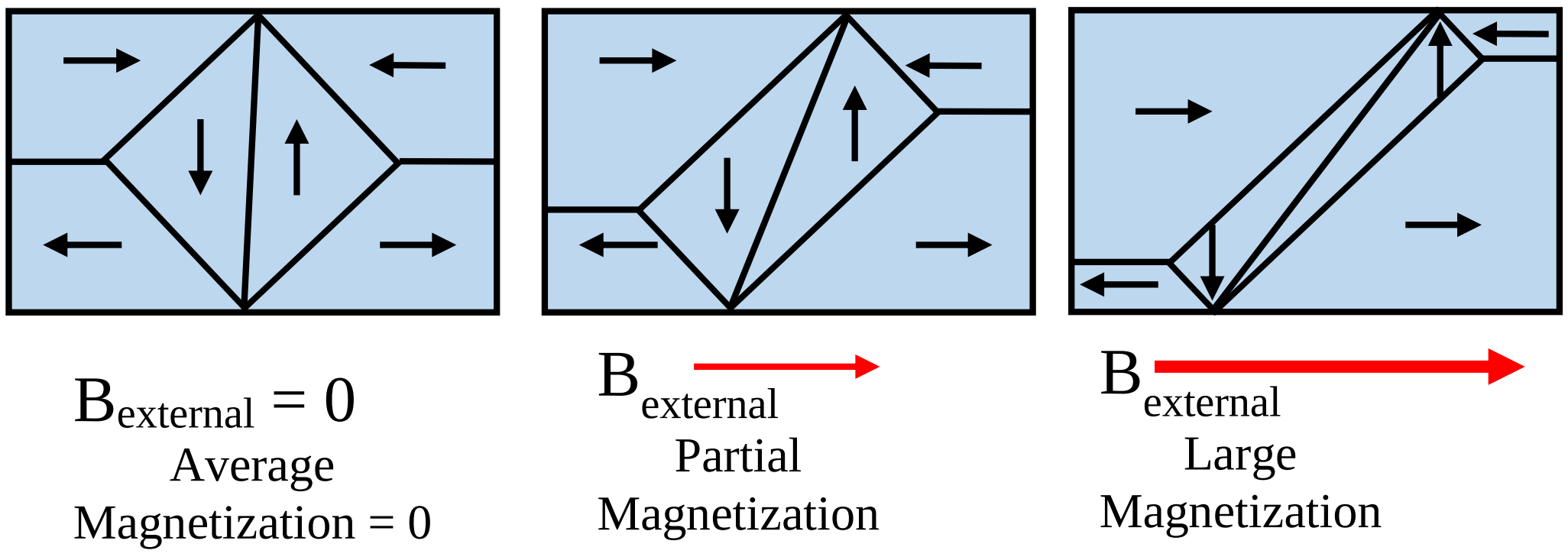

Figure 1. A Conceptual Model for Slowly Varying Domain Boundary Re-Arrangement

## Metglas 2605SA1 Properties

Metglas is an amorphous metallic glass. This means, in principle, that it has no crystalline structure and no grain boundaries. Magnetic domain boundaries in Metglas are therefore free of any influence from grain boundaries or crystal lattice orientation. Their orientation is only constrained by the physical boundaries of the material and by the requirement for minimization of energy.

The material safety data sheet for Metglas 2605SA1 lists the following composition: iron 85 -95%, silicon 5-10% and boron 1-5%. By comparison, Metglas 2605 is reported to be 80% iron and 20% boron while Metglas 2605SC is 81% iron, 13.5% boron, 3.5% silicon and 2% carbon.

The advertised properties for two different alloys, Metglas® 2605SA1 & 2605HB1M**,** can be found on the Metglas website [4] and [5]. The maximum DC relative permeability for Metglas 2605SA1 is stated to be in the range of 45,000 to 600,000, depending on heat treatment.

The Metglas of this study was purchased in the form of a long roll, of 20 cm width, and 23 micrometers thickness. The material was not heat treated after it was manufactured as a thin sheet and appears to have extremely high yield strength, so high, in fact, that formability can be problematic.

## Sample Measurements

A cylindrical sample of the material was made with the dimensions listed in Table 1.

| Geometry | cylindrical |
|---|---|
| Number of layers | 1 |
| Material thickness | 23 micrometers |
| Diameter | 3.33 cm |
| Length | 20 cm |
| Overlap | 3 mm |

Table 1 The Geometry of the Sample

The experimental arrangement is shown in Figure 2. The cylindrical sample was located with its axis horizontal within the active volume of a three-axis set of Helmholtz coils. The horizontal components of the Earth's ambient field were cancelled using two pairs of coils, leaving a 480 milliGauss vertical field component transverse to the cylinder axis. A single axis anisotropic magnetic resistance sensor was located at the center point of the cylinder. The orientation of the sensor was vertical. The sensor and its readout electronics were nominally accurate to ±0.5 milliGauss [6].

The overlap joint in the cylinder was oriented at the top in order to minimize perturbation of the attenuated field by leakage through the joint, as per reference [7]. The external field was measured with the cylinder removed from the test area. To measure the field attenuated by the cylinder, the cylinder was placed over the sensor with no other changes to the system. Prior to measurement, the cylinder was carefully demagnetized to the extent possible, given the 550 mGauss ambient local field.

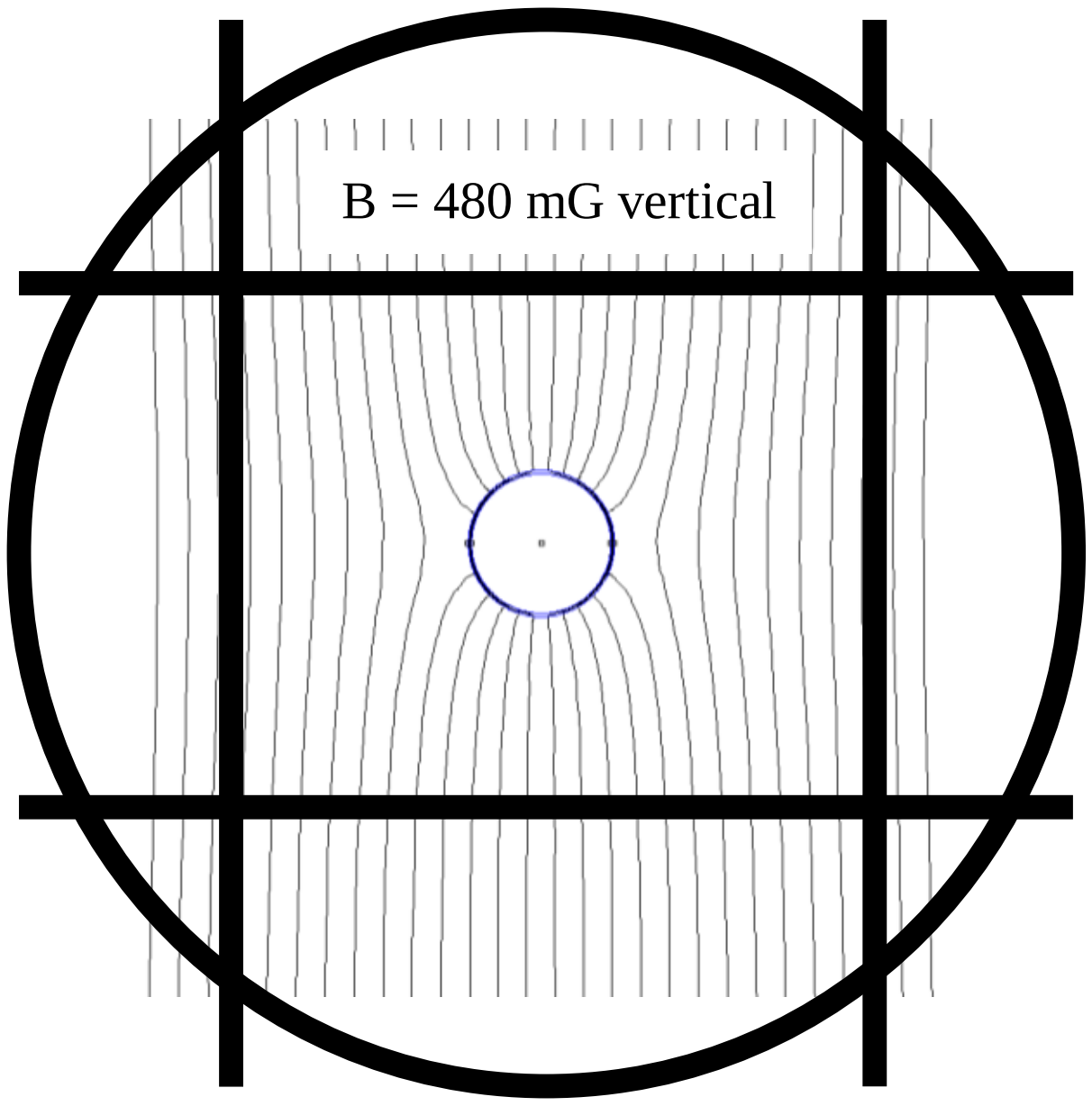

Figure 2. The Arrangement for Measurement of the Attenuated Field

All measurements reported here were made at a room temperature of approximately 20C ±2C. A typical measured relaxation curve for the cylinder is shown in Figure 3.

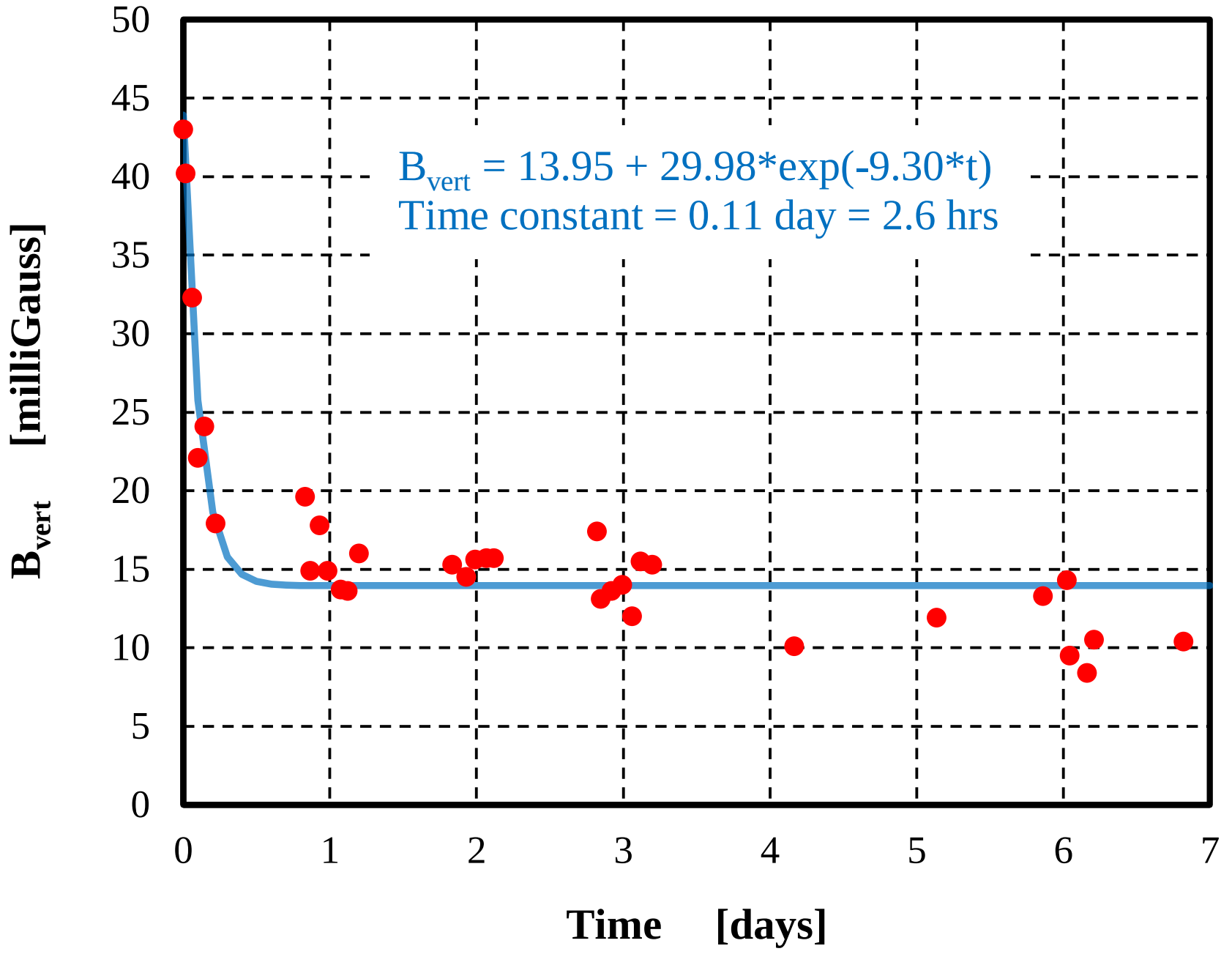

Figure 3. Time Variation of the Magnetic Field Attenuated By the Cylinder

Time variation of the attenuated field was found to be repeatable. After seven days, the cylinder was rotated about its axis through an angle of 180 degrees. The measured field reverted to 43 milliGauss and an identical relaxation process progressed.

Apparently random fluctuations in vertical field readings inside the cylinder were observed at the level of ±3 milliGauss. The fluctuations appear as scatter of plotted points in Figure 3. These fluctuations occurred over time periods of hours and were initially thought to be the result of changes in the local magnetic environment. However, when the Metglas cylinder was not in place, the magnetic field was constant to within ±0.2 milliGauss over a period of one week. This makes it clear that the variation of the local magnetic environment, including variation of the cancellation coil applied field, is not responsible for the scatter. It appears to be due to small, random changes in the magnetization of the Metglas. What drives the change in a random manner? Is it related to the thinness of the magnetic layer? One millimeter thick samples of polycrystalline nickel-iron shielding material in the same experimental arrangement do not exhibit this effect. The physical source of this instability is not understood, and is a worthy subject for further study.

## Analysis

Figure 3 shows the exponential function that best describes the initial drop in the field inside the cylinder. The fit to the data points in Figure 3 was done with a nonlinear least squares regression in three variable parameters. The exponential function is a simplistic way of providing a useful figure of merit, the time constant, to describe the gross characteristics of the first few hours of the field decay curve. It is clearly not a good fit to the data or a good model for the behavior of the cylindrical geometry after one day has elapsed. An argument for why this is the case is as follows: relaxation properties of Metglas are likely to be a function of the local magnetic field magnitude within each unit volume of the material. Lower field is associated with longer relaxation times. In the cylindrical shield geometry, magnetic field is highly non-uniform. A field map of a cylinder with high permeability immersed in a 480 milliGauss ambient field is shown in Figure 4. The wall thickness used for the cylinder in Figure 4 has been greatly increased over that of the Metglas cylinder in order to illustrate the behavior of field lines within the material. The characteristic pattern of flux concentration at the sides of the cylinder can be clearly seen.

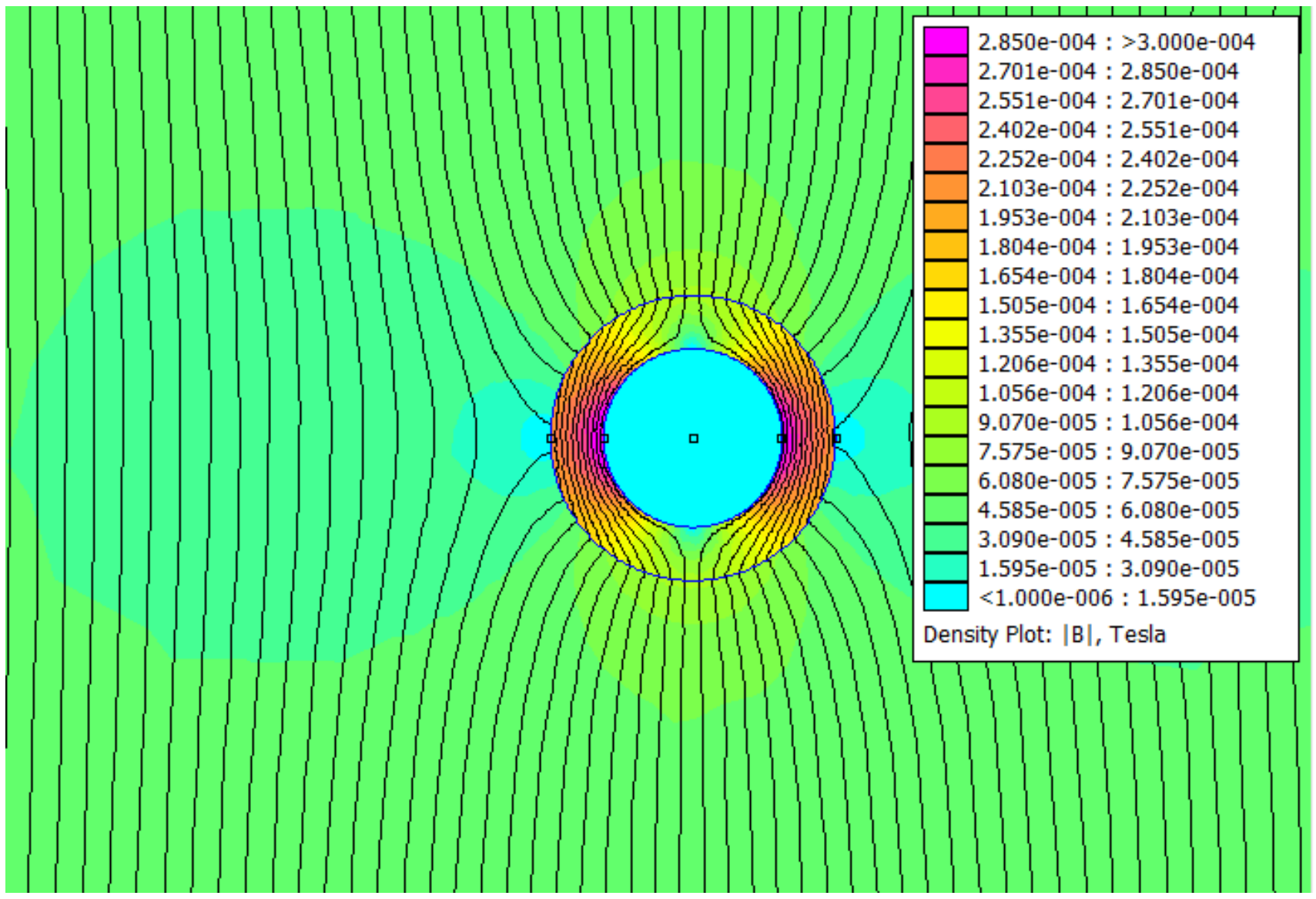

Figure 4. Non-uniformity of Magnetic Field Inside the Material of a Cylindrical Shield

For the thin walled Metglas cylinder, after an initial drop with a time constant of 2.6 hours, the attenuated field slowly decreases to a value between 10 mGauss and 15 mGauss after four days. This allows us to estimate an effective final relative permeability for the Metglas in the relaxed state. The term "effective permeability" is used to denote permeability as calculated from the measured attenuation of the cylinder. Effective permeability is the result of averaging over many factors that may differ between separate locations in the cylinder.

**Method 1 Traditional Analytical Formulary**

Assuming a constant value for the permeability of the Metglas allows use of a typical formula for single layer cylindrical shields [8]:

$$A = \mu_r \cdot \frac{d}{D} + 1 \qquad \text{Equation 1.}$$

A is the transverse field attenuation factor, d is the shield thickness, D is the shield diameter, and $\mu_r$ is the relative magnetic permeability. Using the upper and lower limits to the relaxed field value results in two values for attenuation: $A_{10mG}$ = 480 mG /10 mG = 48 and $A_{15mG}$ = 480 mG /15 mG = 32. The estimated relaxed relative permeability, using the formula above and dimensions from Table 1, is therefore in the range:

$$\mu_r = 57{,}000 \pm 12{,}000$$

**Method 2 Finite Element Model Using a Non-Linear B-H Curve**

The low field B-H curve for Metglas 2606SC was used as a substitute for the Metglas 2606SA1 curve to model non-linearity effects. The appropriate 2605SA1 curve was not readily available, and the gross features of the two materials are substantially the same. The curve used comes from reference [9] and results in very good agreement with the average measured transverse attenuation factor of 40. The B-H curve is shown in Figure 5.

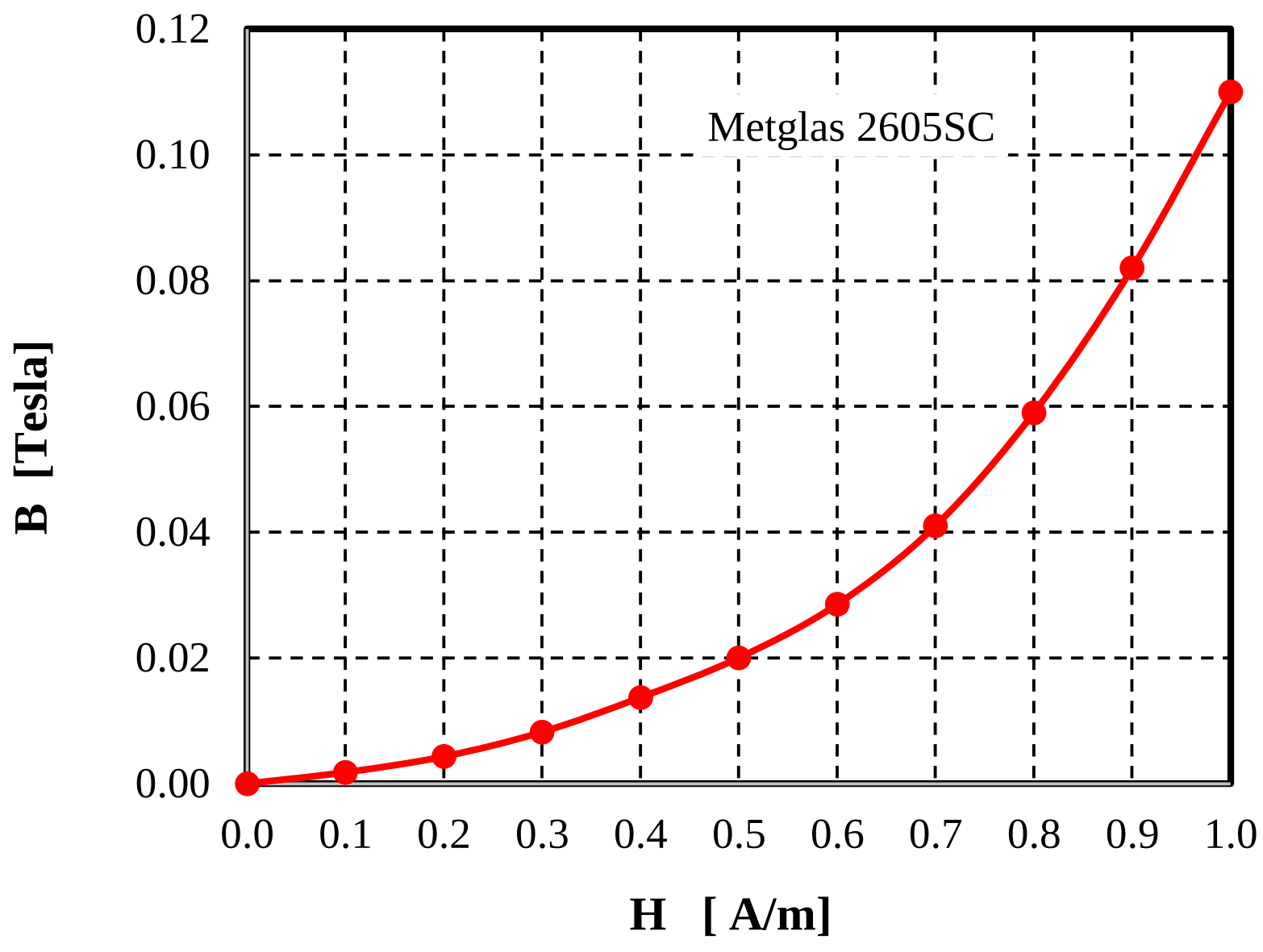

Figure 5. Metglas 2605SC Initial B-H Curve

Results of the finite element calculation are shown in Table 2. The program used was FEMM, V4.2 [10].

| Parameter | Value | Units |
|---|---|---|
| External Field | $4.8 \times 10^{-5}$ | Tesla |
| Transverse Attenuation Factor | 40 | |
| Minimum Relative Permeability | 14,000 | |
| Maximum Relative Permeability | 64,000 | |
| Maximum Field in the Metglas | $6.75 \times 10^{-2}$ | Tesla |

Table 2. Values from the Finite Element Model (Fitted Values are in Blue)

The minimum of permeability occurs at the 12 and 6 o'clock positions on the cylinder. The permeability increases with flux concentration as the 3 and 9 o'clock positions are approached as per the assumed B-H curve.

The attenuated field measurements from Figure 3 can be converted to calculated effective relative permeability, as shown in Figure 6, by use of equation 1.

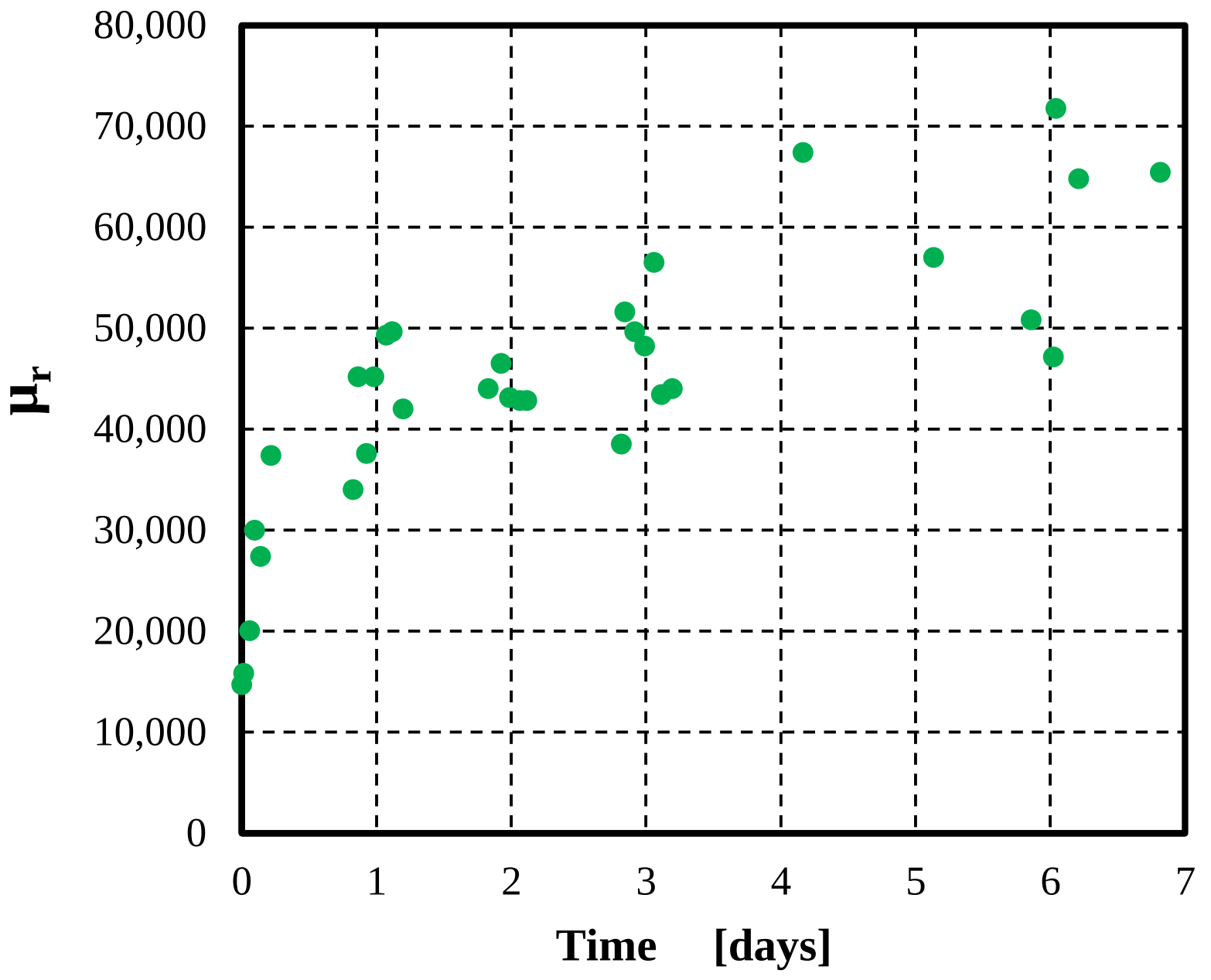

Figure 6. Effective Permeability as a Function of Time

**Conclusion**

Profound magnetic relaxation has been observed in Metglas 2605SA1. Realization of maximum permeability requires a waiting period of approximately four days at room temperature after each change in the magnetic environment of the cylindrical shield.

Two presented methods of approximating the effective relaxed relative permeability of are in relatively good agreement, allowing a value of 57,000 to be used for the case of simple estimation based on constant permeability and immersion of a 3.3 cm diameter by 23 micrometer thick cylindrical shield in the Earth magnetic field. For applied field of 1 Ampere/meter, or less, the initial B-H curve for Metglas 2605SC can be substituted for Metglas 2605SA1 in non-linear models.

Non-uniform flux concentration in the cylindrical geometry means that magnetic relaxation and effective permeability will depend on the diameter and thickness of a particular shield geometry as well as on the magnitude of the local ambient magnetic field.

**Open Questions**

Could the relaxation process be assisted? It has been known for centuries that if one beats on a piece of steel, for example, a sword blade, while keeping it at a constant orientation with respect to the Earth field, it will become magnetized to the extent that it is capable of repelling one pole of a compass needle. What if, rather than beat on the Metglas cylinder, decreasing amplitude ultrasonic wave energy was used to stimulate grain boundary re-arrangement? Could such a technique be safely applied to a complete, in-situ, superconducting RF cryomodule?

## References

[1] Wu, G., et al., “Magnetic Foils for SRF Cryomodule”, SRF2015
http://srf2015proc.triumf.ca/prepress/papers/tupb099.pdf

[2] Galenko, P. P. and Branovitskaya, T. A., “Magnetic Relaxation of Electrotechnical Steel in Constant Magnetizing Fields at Various Temperatures”, Inzhenerno-Fizicheskii Zhurnal, Vol. 46, No. 4, pp. 650–654, April, 1984 http://link.springer.com/article/10.1007%2FBF00826410#page-1

[3] P. Weiss (1906) La variation du ferromagnetisme avec la temperature, *Comptes Rendus*, **143**, p.1136-1149
https://archive.org/stream/ComptesRendusAcademieDesSciences0143/ComptesRendusAcadmieDesSciences-Tome143-Juillet-dcembre1906#page/n1135/mode/2up
as referenced in: Cullity, B., “Introduction to Magnetic Materials”,1972, p.119.

[4] http://www.metglas.com/assets/pdf/2605sa1.pdf

[5] http://www.metglas.com/products/magnetic_materials/2605sa1.asp

[6] https://www.trifield.com/content/dc-milligauss-meter/

[7] Crawford, A., “Modeling of Mechanical Overlap Joints in Magnetic Shields”,
http://arxiv.org/ftp/arxiv/papers/1510/1510.03846.pdf

[8] Tesla Test Facility Conceptual Design Report, 1995, p. 162,
http://tesla.desy.de/TTF_Report/CDR/pdf/cdr_chap4.pdf

[9] Hernandez-Jimenez, G. and Valenzuela, R., “The Time Treatment Effect on the Magnetic properties of Metglas Amorphous Ribbons”, Figure 3, p280 of Moran-Lopez, J., New Trends in Magnetism, Magnetic Materials, and Their Applications, 1994.

[10] Meeker, D., Finite Element Method Magnetics, Version 4.2 (15Nov2013 Build), http://www.femm.info